\documentclass[twocolumn,showpacs,preprintnumbers,amsmath,amssymb]{revtex4}
\usepackage{graphicx}% Include figure files
\usepackage{dcolumn}% Align table columns on decimal point
\usepackage{bm}% bold math

\begin{document}

\preprint{APS/123-QED}

\title{Pressure induced magnetic ordering in Yb$_2$Pd$_2$Sn \\
with two quantum critical points}% Force line breaks with \\

\author{T. Muramatsu$^{1}$}
\author{T. Kanemasa$^{1}$}%
\author{E. Bauer$^{2}$}%
\author{M. Giovannini$^{3}$}%
\author{T. Kagayama$^{1}$}%
\author{K. Shimizu$^{1}$}% \\

\affiliation{%
$^{1}$Kyokugen, Center for Quantum Science and Technology under Extreme Conditions, 
Osaka University, Toyonaka, Osaka 560-8531\\}%
\affiliation{%
$^{2}$Institute of Solid State Physics, Wiender Hauptstrasse 8-10, Vienna University of Technology, Wien A-1041, Austria\\}%
\affiliation{%
$^{3}$Dipartimento di Chimica e Chimica Industriale, University of Genova, Via Dodecaneso 31,?I-16146?Genova
\\}%
\date{\today}% It is always \today, today,

\begin{abstract}
Pressure induced long range antiferromagnetic order is discovered in Yb$_2$Pd$_2$Sn by measuring the electrical resistivity under pressure up to 5.0 GPa. 
Magnetic ordering is observed above about 1.0 GPa, being the lowest pressure in Yb-intermetallics showing pressure induced magnetic ordering. 
Unexpectedly, ordering disappears above about 4.0 GPa, 
giving rise to the first observation of the appearance of two quantum critical points persisting in a broad range of pressure within a single material.
\end{abstract}

\pacs{Valid PACS appear here}% PACS, the Physics and Astronomy
%\keywords{}%Use showkeys class option if keyword
\maketitle

A critical temperature where some kind of ordered state changes to disorder can be controlled by several parameters, 
e.g., pressure, chemical substitution and magnetic field, 
and the point where the critical temperature is tuned to absolute zero temperature is called a quantum critical point (QCP). 
At the QCP, quantum fluctuations develop, inducing a variety of unusual electronic properties. 
Many investigations have already been carried out to understand quantum criticality in strongly correlated electron systems.

Among them, several Ce based intermetallic compounds were studied under applied pressure and magnetic fields
 in order to analyze the properties right at the QCP.
 In general, the magnetic phase transition of Ce compounds can be tuned towards the QCP by the application of external pressure.
 In the proximity of the QCP, strong deviations from Landau Fermi liquid behavior due to the low energy and extended spin fluctuations are observed. 
Some of such systems, e.g., CeCu$_2$Ge$_2$[1], CeRhIn$_5$[2], CePt$_3$Si[3] and CeRhSi$_3$[4], show even exotic superconductivity 
with the possibility of a magnetically mediated pairing mechanism, enhanced at the QCP.

This suppression of the magnetic transition temperature in Ce-intermetallic compounds by applying pressure is well described by a competition between RKKY and Kondo interaction. 
In contrast, the application of pressure to Yb-based intermetallic compounds may induce the magnetically ordered phase. 
Examples here are e.g. YbCu$_2$Si$_2$[5,6], YbNi$_2$Ge$_2$[7], Yb$_2$Ni$_2$Al[8] and YbCuAl[9].
The reason of which can be explained by a change of the valence, 
driving  divalent non-magnetic Yb$^{2+}$(4f$^{14}$) towards the trivalent magnetic Yb$^{3+}$(4f$^{13}$) state owing to a reduction of the unit cell volume. 
Note, the volume associated with the 4f$^{14}$ configuration is larger that that of magnetic 4f$^{13}$. 
Quantum criticality on antiferromagnetically ordered YbRh$_2$Si$_2$ , with a very low ordering temperature $T_\mathrm{N}$  = 65 mK, was well studied by precise magnetic field dependent measurements [10]. 
In general, however, the critical pressure to accomplish quantum criticality in most of the Yb-compounds investigated is located at quite high pressure, above 5 GPa [5,6,7,8,9]. 
For that reason, precise properties at the QCP in Yb compounds are still an open question, 
because probes satisfying the experimental requirements for both low temperature measurements (below 1 K) and high pressure (above 5 GPa) are quite limited to a few methods, 
e.g., the diamond-anvil-cell (DAC) or a Bridgman type pressure cell assembled to a dilution refrigerator.

Resuming our recent investigations regarding Yb compounds, the present study focuses on Yb$_2$Pd$_2$Sn, one end member of solid solution Yb$_2$Pd$_2$In$_{1-x}$Sn$_x$ [11,12]. 
Yb$_2$Pd$_2$In$_{1-x}$Sn$_x$ crystallizes in the tetragonal Mo$_2$FeB$_2$-type in which alternating layers of Yb and the other one made of Pd and Sn stack sequentially along c-axis. 
With respect to Yb$_2$Pd$_2$In$_{1-x}$Sn$_x$, both border compounds remain non-magnetic, while long range magnetic order appears in a narrow concentration range from about x = 0.6 to x = 0.9 without changing the crystal structure. 
To the best of our knowledge, this constitutes the first example of a single series with the possibility of two QCPs.  
This characteristic feature might be attributed to both chemical pressure onto the Yb ion by substituting In by Sn having different ionic radii, and the non-isoelectronic character of the In/Sn substitution.

In Ref. [11], the pressure effect on one border compound, Yb$_2$Pd$_2$Sn, of the series Yb$_2$Pd$_2$In$_{1-x}$Sn$_x$ has been investigated by measuring the electrical resistivity, $\rho$, up to 1.7 GPa. 
$\rho(T)$ of Yb$_2$Pd$_2$Sn exhibits at ambient pressure two broad maxima at around $T^{high}_{max}$ $\sim$ 200 K and about $T^{low}_{max}$ $\sim$ 11 K. 
$T^{low}_{max}$ is concerned with the characteristic temperature where a coherent Kondo effect starts to develop with decreasing temperature, 
while $T^{high}_{max}$ is caused by incoherent Kondo scattering under the influence of crystalline electric field (CEF) splitting of the 4f state of the Yb ion. 
Both features are typically observed in Yb and Ce intermetallics. 
However, no signs of long range magnetic order were observed within the measurement range (1.5 K $\leq T \leq$ 300 K and 0 GPa $\leq P \leq$ 1.7 GPa) studied. 
The aim of the present paper is searching for new QCPs located within a pressure region, accessible for more experimental methods. 
Attempting to observe pressure induced magnetic ordering in Yb$_2$Pd$_2$Sn, we measured the temperature dependent electrical resistivity $\rho(T)$ at higher pressures, above 1.7 GPa, and at lower temperatures, below 1.5 K.

\begin{figure}[tb]
%\begin{center}
\includegraphics[width=8.0cm]{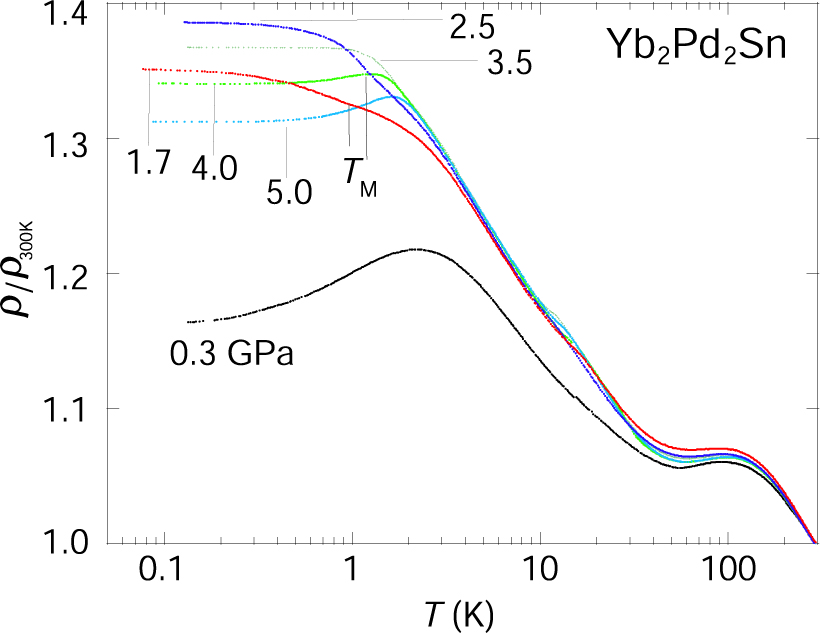}
%\end{center}
\caption{
(color online) 
Temperature dependence of the electrical resistivity normalized by the resistivity at 300 K of Yb$_2$Pd$_2$Sn at various pressures. 
Magnetic transition temperatures $T_\mathrm{M}$ are indicated by arrows. 
The anomalies observed between 10 and 20 K are not intrinsic 
and occur due to the fluctuation of the inner pressure of clamped-type pressure cell 
through temperature change.
}
\label{f1}
\end{figure}
A polycrystalline sample of Yb$_2$Pd$_2$Sn was prepared from stoichiometric amounts of elements by high frequency melting the constituent materials in a closed tantalum crucible. 
A subsequent heat treatment at 1250 K for one week served to ensure phase purity. 
Synthesized Yb$_2$Pd$_2$Sn crystallizes with the $D^{5}_{4h}$ tetragonal space group. 
The 4h site occupied by Yb has a low mm point symmetry. 
To achieve a wider pressure range with high resolution of resistivity data, 
we employed a DAC and embedded the conductive electrodes into the sample space for a four-probe electrical resistivity measurement. 
The sample and some small ruby chips were clamped in the sample space with NaCl as a pressure transmitting medium. 
Determination of the pressure value was carried out by the ruby fluorescence method at about 20 K.
 Pressure distribution in the sample was within about 10 \% and pressure change from room temperature to 30 K is smaller than 0.5 GPa. 
The DAC was then assembled in the $^3$He/$^4$He dilution refrigerator, and the sample was cooled down to 70 mK. 
Other details of the DAC experiment are given in ref. 13.

\begin{figure}[tb]
%\begin{center}
\includegraphics[width=8.5cm]{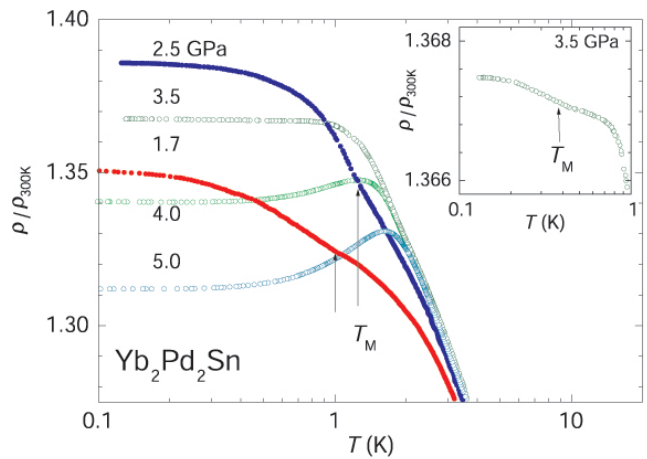}
%\end{center}
\caption{
(color online) 
Close-up plot of temperature dependence of the electical resistivity normalized at 300 K near the magnetic ordering temperature of Yb$_2$Pd$_2$Sn.
 Inset shows the anomaly observed at 3.5 GPa. 
}
\label{f1}
\end{figure}
Fig.1.shows $\rho(T)$ normalized by the resistivity at 300 K at several pressures. 
Overall feature of the $\rho(T)$/$\rho(300K)$ is characterized by a negative logarithmic behavior 
which originates from the dominant contribution of magnetic scattering to the resistivity in Yb$_2$Pd$_2$Sn. 
At lowest pressure of 0.3 GPa, two broad maxima are observed as often seen in heavy fermion compounds. 
The broad low temperature maximum $T^{low}_{max}$ reflects coherent Kondo interaction and it is regarded as a measure of the Kondo temperature $T_\mathrm{K}$ [14]. 
The broad high temperature maximum $T^{high}_{max}$ is caused by the incoherent Kondo effect under the influence of CEF splitting of the 4f$^{13}$-electronic state.

At 1.0 K and 1.1 K in 1.7 GPa and 2.5 GPa, respectively, we found a kink-like anomaly, indicated by arrows in Fig.1 and Fig.2. 
Also at 3.5 GPa, a tiny kink at 0.4 K is observed (inset of Fig.2), disappearing under a small magnetic field of 0.5 T. 
The overall shapes of $\rho(T)$ at 1.7 GPa and 2.5 GPa are rather similar to that of Yb$_2$Pd$_2$In$_{0.2}$Sn$_{0.8}$, [11] 
where the occurrence of bulk antiferromagnetic order is confirmed by various measurements. 
Therefore, we conclude that the anomaly observed reflects the onset temperature $T_\mathrm{M}$ of magnetic order, 
triggered by the emergence of the full trivalent state of Yb under compression of the unit cell volume. 
The enhancement of the resistivity below $T_\mathrm{M}$ can be explained in terms of a superzone boundary effect owing to the quasi-2D crystal structure of Yb$_2$Pd$_2$In$_{1-x}$Sn$_x$. 
Below $T_\mathrm{M}$, the electrical resistivity rises due to the decrement of the 4f hole career density as a consequence of nesting at the Fermi energy $E_\mathrm{F}$. 
The ambiguity at $T_\mathrm{M}$ is presumably attributed to the unexpected rapid change of $T_\mathrm{M}$ by pressure and/or the inevitable pressure distribution in the sample.

\begin{figure}[tb]
%\begin{center}
\includegraphics[width=7.5cm]{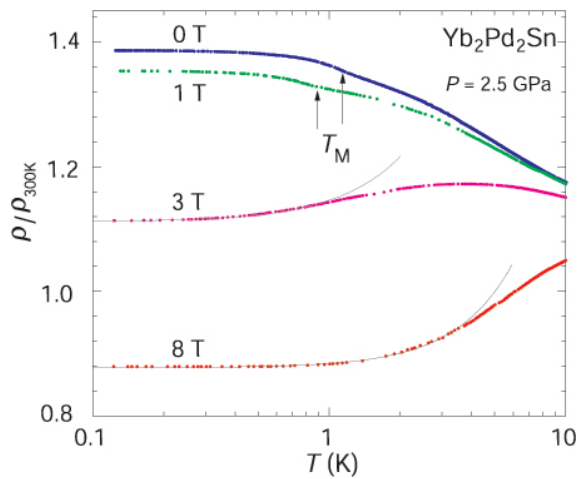}
%\end{center}
\caption{
(color online) 
Temperature dependence of the electrical resistivity normalized at 300 K at 2.5 GPa under magnetic fields. 
Magnetic transition temperatures are indicated by arrows.
Solid lines are fitting results by equation $\rho = \rho_\mathrm{0} + AT^{n}$.
See text for details.
}
\label{f1}
\end{figure}
The field response of $\rho(T)$ at 2.5 GPa for $H \perp I$ is shown in Fig.3. 
Yb$_2$Pd$_2$Sn exhibits a negative magnetoresistance at least for fields up to 8.0 T and the total reduction of resistivity is about 40 \% at the lowest temperature. 
$T_\mathrm{M}$ decreases from 1.1 K to 0.9 K at 0 T and 1.0 T, respectively,
 and the further application of magnetic fields causes the anomaly to disappear, at least above 120 mK. 
This rapid suppression and suppression of $T_\mathrm{M}$ by magnetic fields can be described as an almost unambiguous signature of antiferromagnetic ordering. 
The $\rho(T)$ at 3.0 T obeys the power law of $\rho = \rho_\mathrm{0} + T^{1.6}$, indicating a field derived non-Fermi liquid state. 
Further applying magnetic fields up to 8.0 T, the exponent of  the power law below 3 K increases to 1.8. 
This might refer to a crossover from a non-Fermi liquid to a field induced Fermi liquid ground state, which would follow the equation $\rho = \rho_\mathrm{0} + T^2$.

\begin{figure}[tb]
%\begin{center}
\includegraphics[width=7.5cm]{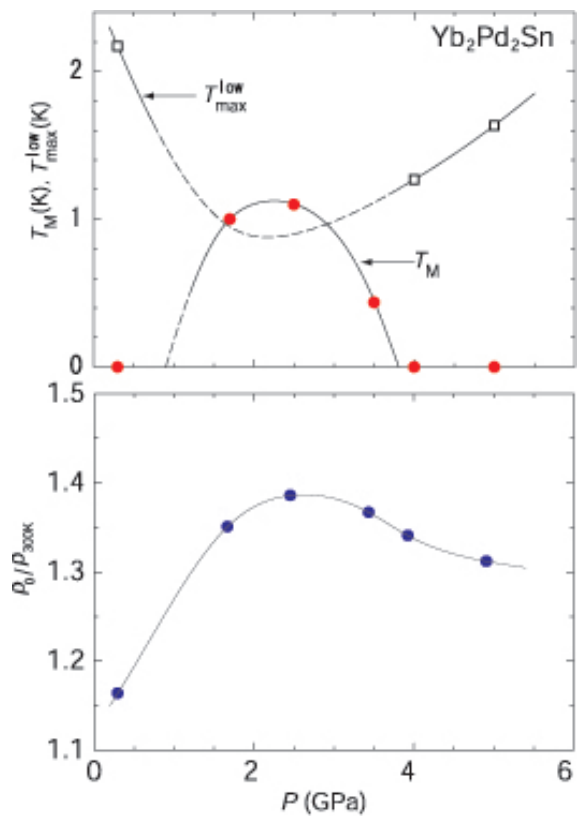}
%\end{center}
\caption{
(color online) 
(a) Pressure-temperature phase-diagram of magnetic ordering temperature and the low temperature maximum of Yb$_2$Pd$_2$Sn. 
Solid and dotted lines are drawn for eye-guide. $T^{low}_{max}$ is not observed at 1.7, 2.5 and 3.5 GPa. 
However, the line passes through below $T_\mathrm{M}$, because $T^{low}_{max}$'s are not seen at least above $T_\mathrm{M}$. 
(b) Pressure dependence of the residual resistivity normalized by the resistivity at 300 K. Solid line is for eye guide.
}
\label{f1}
\end{figure}
Fig.4(a) illustrates the pressure-temperature phase diagram of Yb$_2$Pd$_2$Sn obtained from $T_\mathrm{M}$ and $T^{low}_{max}$ of the resistivity measurement. 
As $T_\mathrm{M}$ is not observed at 0.3 GPa but at 1.7 GPa, the magnetic ordering is induced at around 1.0 GPa and it disappears at the pressure slightly below 4.0 GPa. 
In the pressure range from 1.0 to 4.0 GPa, $T^{low}_{max}$ is not observed and the shape of the resistivity curve changes to the shoulder. 
Increasing pressure above 4.0 GPa, the maximum emerges again, presumably due to the loss of magnetic ordering. 
This feature can be explained by the opposite pressure dependence of $T_\mathrm{K}$ and $T_\mathrm{M}$. 
For lower pressure it is expected that $T_\mathrm{K}$ decreases with pressure, as generally observed in Yb-intermetallics, reaching a minimum around 2.5 Gpa, before increasing again above 4.0 GPa. 
Within the magnetically ordered phase $T_\mathrm{K}$ is not estimated. 
Contrary to this pressure tendency of $T_\mathrm{K}$, $T_\mathrm{M}$ shows a maximum at around 2.0 GPa, 
where $T_\mathrm{M}$ is expected to exceed $T_\mathrm{K}$ This mutual behavior of $T_\mathrm{K}$ and $T_\mathrm{M}$ is quite similar to the solid solution Yb$_2$Pd$_2$In$_{1-x}$Sn$_x$ 
and this might provide corroborative evidence regarding the existence of long range magnetic order. 
Additionally, similar relationships are generally recognized in other studies concerning Yb compounds in which $T_\mathrm{K}$ decreases before saturating towards a maximum pressure in the magnetically ordered phase. 
This feature is also confirmed theoretically [15]. 
The maximum $T_\mathrm{M}$ of about 1.1 K is smaller than that derived by  chemical substitution in Yb$_2$Pd$_2$In$_{1-x}$Sn$_x$. 
This fact might originate from some pressure inhomogneities due to NaCl used as pressure transmitting medium, 
or from differences of $E_\mathrm{F}$ when substituting Sn by In which is expected to lower $E_\mathrm{F}$. 
The suppression of $T_\mathrm{M}$ and associated increase of $T_\mathrm{K}$ above 2.5 GPa is similar to the pressure effect of Ce-based intermetallic compounds, 
i.e., the suppression of $T_\mathrm{M}$ might occur due to a competition between RKKY interaction and Kondo effect, as generally observed in Ce intermetallics. 
The pressures where either magnetic ordering appears or vanishes at 0 K corresponds to a QCP and any other Ce or Yb based intermetallic compounds with two QCPs have not been observed so far. 
Therefore, to the best of our knowledge, this is the first finding of a system with two QCPs originated by magnetic ordering in a heavy fermion system. 
In addition, it is surprising that both QCPs occur at a comparable low pressure region below, 4.0 GPa. 

In Fig.4(b), the pressure dependence of the residual resistivity normalized by the resistivity at 300K, $\rho_\mathrm{0}^\mathrm{n}=\rho_\mathrm{0}/\rho_\mathrm{300K}$, is plotted. 
Obviously, $T_\mathrm{M}$ and $\rho_\mathrm{0}^\mathrm{n}$ match each other as a function of pressure. 
Particularly, $\rho_\mathrm{0}^\mathrm{n}$ takes larger values where magnetic ordering occurs. 
In general, $\rho_\mathrm{0}$ can be expressed by $\rho_\mathrm{0} = (\hbar/e^2l)(3\pi^2)^{1/3}n^{-2/3}$, where $l$ is the mean free path of carriers and $n$ is carrier density. 
From this equation it follows that the enhancement of $\rho_\mathrm{0}^\mathrm{n}$ at about 2.0 GPa refers to either $l$ or $n$ must be reduced. 
The mean free path at low temperature limit, however, could not be affected by pressure, 
since it may be determined by the imperfection of crystal lattice. 
The maximum of $\rho_\mathrm{0}^\mathrm{n}$ might thus be attributed to a decrement of the carrier density. 
Such a change of the carrier density would be in line with the occurrence of antiferromagentic ordering, 
since there a partial opening of an energy gap at $E_\mathrm{F}$ should emerge. 
However, it is also known that quantum fluctuations such as magnetic fluctuations increase $\rho_\mathrm{0}$ at QCP as well [16,17]. 
That must be examined by further measurement in future experiments.
   
In summary, we found pressure induced magnetic ordering in Yb$_2$Pd$_2$Sn. 
Magnetic ordering is confirmed by the response of $T_\mathrm{M}$ to magnetic fields and the pressure dependence of a specific relationship between $T_\mathrm{M}$ and $T^{low}_{max}$. 
Coincidence of two maxima of $\rho_\mathrm{0}$ and $T_\mathrm{M}$ also supports the existence of a magnetically ordered state. 
Long range magnetic order exists above about 1.0 GPa but disappears at about 4.0 GPa. 
This indicates that two QCPs exist in the pressure-temperature phase diagram. 
Such a scenario is the first observation that two QCPs are observed within a single compound. 
Moreover, it is notable that they are located at rather low pressures, 
offering the proposition that other Yb compounds also have two QCPs in the low pressure region, which my stimulate further research in this field. 

We thank K. Miyake for many fruitful discussions. 
This work was also supported by MEXT.KAKENHI(15204032) and the 21st century COE Program (G18) of Japan Society for the Promotion of Science. 
T.M. is grateful for financial support by the 21st century COE Program 
and E.B likes to acknowledge the Asutrian FWF, P 18054 and the Osaka University for financial support.

\end{document}